\documentstyle[aps,multicol,epsfig,prl,fancyheadings]{revtex}

\begin{document}
\input epsf
\draft \preprint{HEP/123-qed}
\title{Analysis of the elasto-plastic response of a polygonal packing}
\author{F. Alonso-Marroquin$^*$, H. J. Herrmann$^*$ and S. Luding$^{*,**}$}
\address{(*) ICA1, University of Stuttgart, \\
Pfaffenwaldring 27 \\
70569 Stuttgart, Germany\\}
\address{ (**)Technische Universiteit Delft (TUD), \\
       DelftChemTech, Particle Technology, \\
       Julianalaan 136, 2628 BL Delft, The Netherlands\\}
\date{\today}

%
\maketitle

\begin{abstract}

We investigate the constitutive response of two-dimensional packed samples of polygons
using molecular dynamics simulation. The incremental elasto-plastic response is examined 
in the pre-failure regime. Besides the Young modulus and the Poisson ratio, an additional
parameter must be included, which takes into account the anisotropy of the elastic response. 
The plastic deformations are described by the introduction of the yield and the flow directions. These directions do not agree, which reproduces the non-associated feature of realistic
soils. In order to detect the yield surface, different loading-unloading-reloading tests 
were performed. During the reload path, it is found that the yielding develops continuously 
with the amplitude of loading, which does not allow to identify a purely elastic regime. 

\end{abstract}

\begin{multicols}{2}

\section{Introduction}
\label{Intro}

Traditionally, the quasi-static deformation of soils has been described by using
constitutive laws. They are empirical relations between the stress and the 
strain involving a certain number of material parameters, which, in the simplest models,
can be measured in experimental tests \cite{vermeer,darve,kolymbas}. However, the more 
sophisticated models involve so many parameters that their direct experimental meaning and 
their identification becomes impossible.

In the last years the numerical simulations have been used as an alternative to study 
the behavior of soils. Usually, disks or spheres are used in order to capture the
granularity of the materials \cite{cundall,radjai,thornton,bardet}.  
The simplicity of their geometry allows to reduce the computer time of  calculations. 
However, they do not take into account the diversity of shapes of the grains 
in realistic materials. 

A more detailed description is presented here by using randomly generated 
convex polygons. The interaction between the polygons can be handled by letting the polygons 
interpenetrate each  other and calculating the force  as a function of their overlap 
\cite{tillemans}. This approach has  been successfully applied to model different processes, 
like  fragmentation \cite{Kun,superkun},  damage \cite{Kun2,addetta}, strain localization and 
earthquakes \cite{tillemans}. The contribution of this work is to determine the constitutive
relation of this discrete model material in the regime of quasi-static deformations. The results 
show that simple mechanical laws at the grain level are able to reproduce the complex 
macroscopic behavior of the deformation of soils.

The details of the particle model are presented in Sec.\ \ref{model}. In addition to the 
normal contact force mentioned above, the tangential contact force law 
is implemented by a Coulomb friction criterion, and the boundary conditions are modeled by the 
introduction of a flexible membrane that allows to fix the stress value. 
The calculation  of the constitutive relations is presented 
in Sec.\ \ref{c-r}. We discuss the results in the framework of the classical theory of 
elasto-plasticity. The summary and the perspectives of this work are presented in
Sec.\ \ref{conclusion}.

\section{Model}
\label{model}

The polygons representing the particles of this model are generated using a simple version of the 
Voronoi tessellation:  First, we choose a random point in each cell of a regular square lattice, 
then each polygon is constructed assigning to each point that part of the plane that is nearer 
to it than  to any other point. Each polygon is subjected to interparticle contact forces and 
boundary forces as we explain below. 

When two polygons overlap, two contact points appear form the intersection of their edges.
The contact line is defined by the segment connecting these two intersection points. The contact 
force is calculated as

\begin{equation}
  \vec{f^c}=k_n \Delta x^c_n \hat{n}^c + k_t \Delta x^c_t \hat{t}^c,
\label{c-f}
\end{equation}

\noindent 
here $\hat{n}^c$ and $\hat{t}^c$ denotes the normal and tangential unitary vectors with respect to the
contact line, and $k_n$ and $k_t$ are the stiffness in the respective directions.  
The overlapping length $\Delta x^c_n$  is the ratio between
the overlap area of the polygons and the length of their contact line. 
$\Delta x^c_t$ defines the elastic tangential displacement of the contact, 
that is given by the time integral starting at the begin of the contact

\begin{equation}
\Delta x^c_t=\int_{0}^{t}v^c_t(t')\Theta(f^c_t - \mu f^c_n)dt',
\label{t-d} 
\end{equation}

\noindent
Where $\Theta$ is the Heaviside function and $\vec{v}^c_t$ denotes the tangential component of 
the relative velocity $\vec{v}^c$ at the contact. $\vec{v}^c$ depends on the linear velocity 
$\vec{v}_i$ and angular velocity $\vec{\omega}_i$ of the particles in contact according to

\begin{equation}
\vec{v}^c=\vec{v}_{i}-\vec{v}_{j}-\vec{\omega}_{i}\times\vec{\ell}_{i}
+\vec{\omega}_{j}\times\vec{\ell}_{j}.
\end{equation}

\noindent
The branch vector $\vec{\ell}_i$ connects the center of mass of  particle $i$ with 
the point of application of the contact force. This point is taken as the center of mass of the 
overlapping polygon. Eq.\ (\ref{t-d}) defines a limit of elasticity in the contact force. When 
the contact force satisfies $f^c_t = \mu f^c_n$ the contact slips, giving rise
to a plastic deformation. 

The external forces are applied on the boundary through a flexible membrane which
surrounds the sample. Such a membrane is calculated using an iterative algorithm, that 
selects the segments of the external contour whose bending angle is smaller than a 
threshold angle  $\theta_{th}$ \cite{alonso}. On each selected segment 
$\vec{T}=\Delta x_1 \hat{x}_1+\Delta x_3 \hat{x}_3$, we apply an external force of the form

\begin{equation}
\vec{f}^b=-\sigma_1\Delta x_3 \hat{x}_1 + \sigma_3 \Delta x_1 \hat{x}_3.
\label{fbound}
\end{equation} 

Here $\hat{x}_1$ and $\hat{x}_3$ are the unit vectors of the Cartesian coordinate system.
$\sigma_1$ and $\sigma_3$ are the components of the stress we want to apply on the sample, 
as it is presented in Sec. \ref{c-r}.

\begin{figure}[htb]
\begin{center}
 \epsfig{file=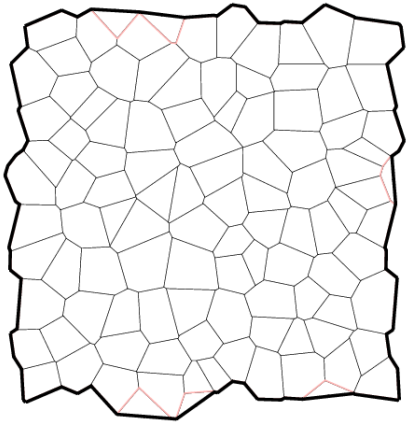,width=7.0cm,angle=0,clip=1}
 \end{center}
\caption{Schematic plot of the membrane obtained with threshold bending angle $\pi/2$.}
\label{memb}
\end{figure}

The contact forces and the boundary forces are inserted in Newton's equations of motion which is 
solved numerically using a predictor-corrector algorithm.  In order to enhance the stability 
of the numerical method and to allow for rapid relaxation, some viscous forces are included both 
in the contacts and in the boundaries:

\begin{eqnarray}
\vec{f}^c_v  &=& -m(\gamma_n v^c_n \hat{n}^c + \gamma_t v^c_t \hat{t}^c), \nonumber\\
\vec{f}^b_v  &=& -m_i \gamma_t \vec{v_i} ~. 
\label{dm}
\end{eqnarray}

The contribution of these forces is almost negligible in the quasi-static regime where 
velocities are small. They are included only to reduce the acoustic waves emitted when 
the system goes from one equilibrium state to the other. 
$m=(1/m_i+1/m_j)^{-1}$ is the effective mass of the  particles in contact, 
and $m_i$ is the
mass of the particle $i$ in contact with the membrane.

There are three characteristic times in the simulation: The relaxation time 
$t_r = 1/\gamma_n$, the loading time $t_0$ and the characteristic period of 
oscillation $t_s = \sqrt{k_n/m_0}$. Here $m_0$ is the mean mass of the polygons. 
This leads to a minimum set of dimensionless parameters, whose selected 
values are shown in the table.

\vspace{0.5cm}
  
\begin{tabular}{||c|c|c||}
\hline
variable  & ratio  & default  value \\ 
\hline 
time of relaxation   & $t_r/t_s $          & $0.1$\\
time of loading      & $t_0/t_s $         & $1250$\\
friction coefficient & $\mu$               & $0.25$\\ 
stiffness ratio      & $k_t/k_n=\gamma_t/\gamma_n$           & $0.33$ \\
bending angle        & $\theta_{th}$       & $\pi/4$\\ \hline
\end{tabular}

\section{Continuous Relations}
\label{c-r}

The characterization of the macroscopic state of a granular material in static equilibrium
is usually given by the Cauchy stress tensor. The derivation of this tensor over
a representative volume \cite{bagi} leads to

\begin{equation}
\sigma_{ij} = \frac{1}{A}\sum_{b} x^b_{i} f^b_{j} ~.
\label{cauchy}
\end{equation}

The sub-scripts $i$ and $j$ in Eq.\ (\ref{cauchy}) denote the components of vectors and tensors.
Here $\vec{x}^b$ is the point of application of the boundary force $\vec{f}^b$. This force
is defined in Eq.\ (\ref{fbound}). $A$ is the area enclosed by the boundary. The sum goes over all 
the boundary forces of the sample.  Inserting Eq.\ (\ref{fbound}) in Eq.\ (\ref{cauchy}) leads to

\begin{equation}
\sigma= \frac{1}{A}\left[ \begin{array}{cc}
                  -\sigma_1\sum_b{x^b\Delta y^b} &  \sigma_3\sum_b{x^b\Delta x^b}   \\
                  -\sigma_1\sum_b{y^b\Delta y^b}  & \sigma_3\sum_b{y^b\Delta x^b}   
                 \end{array} \right].
\end{equation}

\noindent
These sums can be converted into integrals over closed loops. 
Then, the calculation of such integrals leads to

\begin{equation}
   \sigma        =  \left[ \begin{array}{cc}
                    \sigma_1& 0  \\
                    0  & \sigma_3  
                     \end{array} \right].
\label{stress1}
\end{equation}

\noindent
Thus, the stress eigensystem coincides with the Cartesian
coordinate system used.  We can reduce the notation introducing the {\it pressure} $p$ 
and the {\it shear stress} $q$ in the components of the {\it stress vector}

\begin{equation}
\tilde{\sigma}=\left[ \begin{array}{c}  p\\q \end{array} \right]  
= \frac{1}{2} \left[ \begin{array}{c} \sigma_1+\sigma_3 \\
                                      \sigma_1-\sigma_3  \end{array} \right].
\label{stv}
\end{equation}

In the same way, the incremental strain tensor can be calculated as the average of the 
displacement gradient over the area of the sample. It has been shown \cite{kruyt} that 
this average can be transformed into a sum over boundary segments of the sample  

\begin{equation}
d \epsilon _{ij} = \frac{1}{A}\sum_{b} \Delta x^b_{i} N^b_{j}.
\end{equation}

\noindent
Here ${N}^b$ is the $90^o$ counterclockwise rotation of the boundary segment $\vec{T}$. 
The displacement of the segment $\Delta \vec{x}^b$  is calculated
from the linear displacement $\Delta \vec{x}$ and the angular rotation $\Delta\vec{\phi}$ of the 
polygon, according to

\begin{equation}
\Delta \vec{x}^b = \Delta \vec{x} + \Delta \vec{\phi} \times \vec{\ell}. 
\end{equation}

\noindent

The vector $\vec{\ell}$ connects the center of the segment with the center of mass of the
polygon to which it belongs. 

The eigenvalues  $d\epsilon_1$, $d\epsilon_3$ of the symmetric part of $d \epsilon_{ij}$  define the 
{\it volumetric } and {\it shear}  component of the strain as the components of the 
{\it incremental strain vector}:

\begin{equation}
d\tilde{\epsilon}=\left[ \begin{array}{c} d\epsilon_v\\d\epsilon_{\gamma} \end{array} \right]
  = \left[ \begin{array}{c} d\epsilon_1+d\epsilon_3 \\
                                       d\epsilon_1-d\epsilon_3  \end{array} \right].     
\label{strain}
\end{equation}

From the {\bf macro-mechanic} point of view, each state of the sample is related to a single point 
in the stress space, and the quasi-static  evolution of the system is represented by the movement 
of this point in the stress space.  The
resulting deformation during the transition from stress state $\tilde{\sigma}$  to  
$\tilde{\sigma}+d\tilde{\sigma}$ is given by the incremental strain  $d\tilde{\epsilon}$.
In advance, let us separate the incremental stress in its  {\it elastic} (recoverable) and  
{\it plastic} (irrecoverable) components:

\begin{equation}
d\tilde{\epsilon}= d\tilde{\epsilon}^e+ d\tilde{\epsilon}^p.
\label{elastoplastic}
\end{equation}

\noindent

\begin{figure}[htb]
 \begin{center}
 \epsfig{file=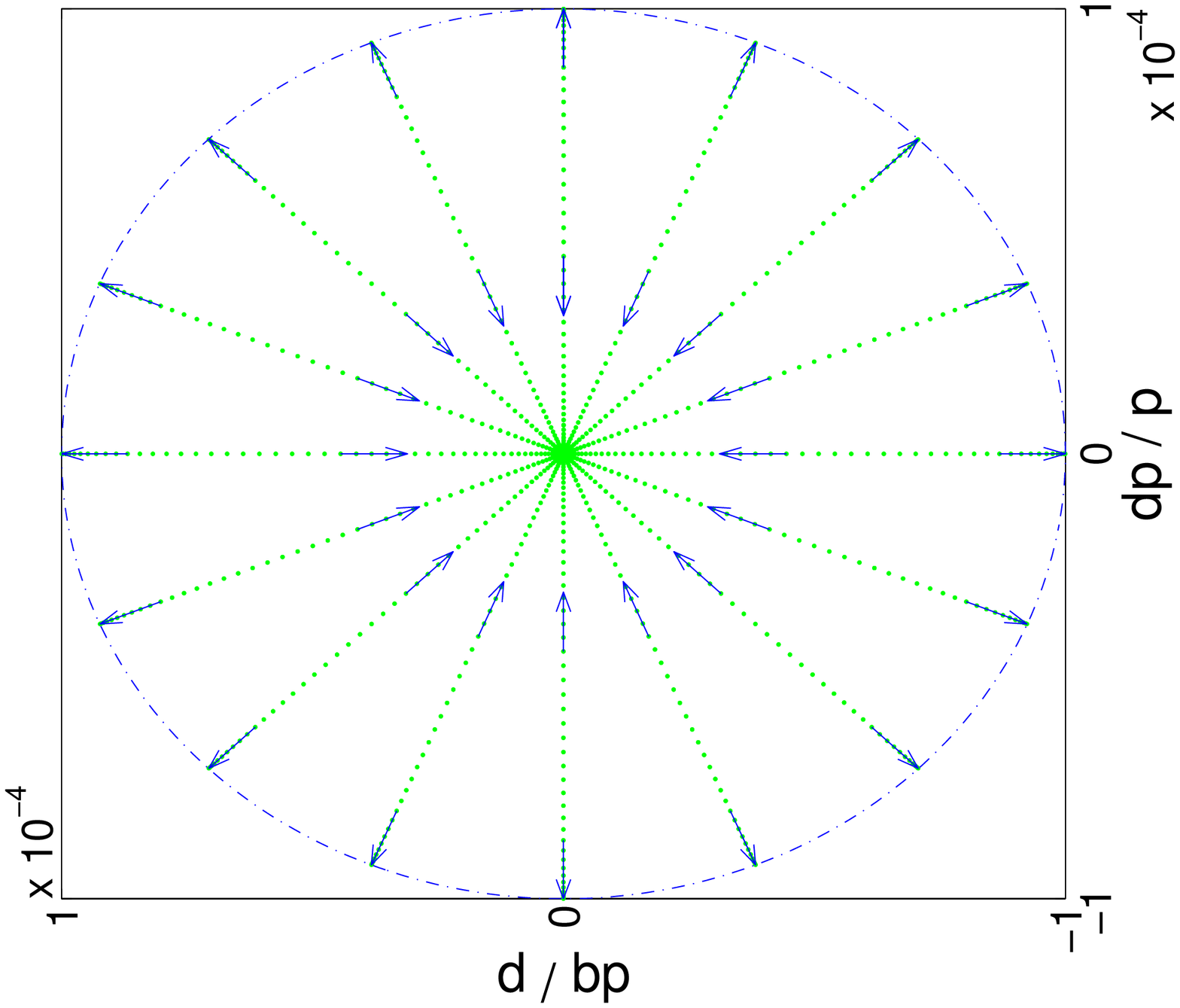,width=6.2cm,height=6.2cm,angle=-90,clip=1}
 \epsfig{file=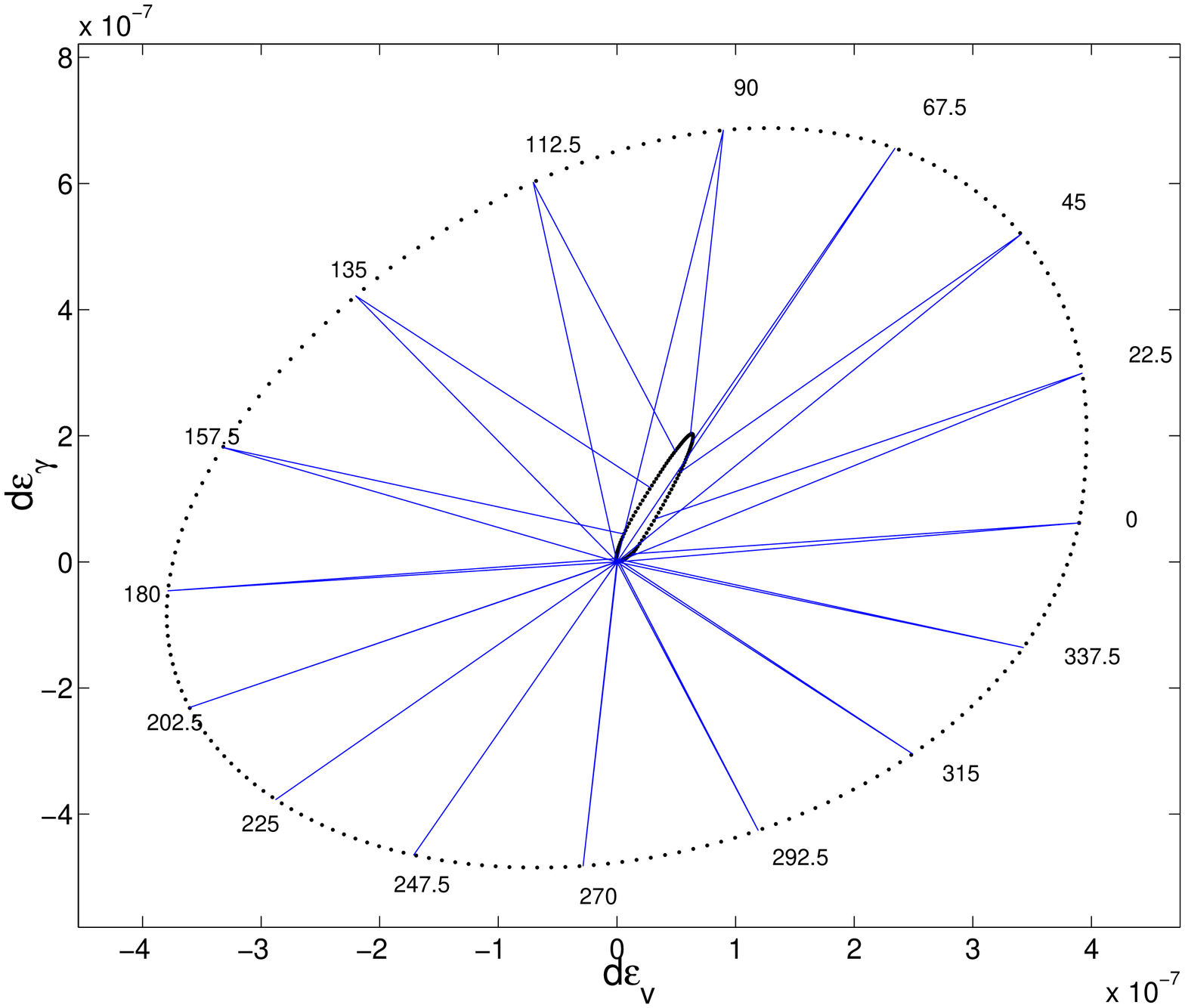,width=9.0cm,height=6.0cm,angle=-90,clip=1}
 \end{center}
\caption{Stress- and strain-relations resulting from the load-unload test. grey lines
represent the paths in the stress and strain spaces. The dotted line gives the strain 
envelope response and the solid line is the plastic envelope response.}
\label{de1}
\end{figure}

Following the procedure proposed by Bardet \cite{bardet} both components can be obtained 
as it is shown in Fig.\ \ref{de1}. Initially, the sample is 
in the stress state $\tilde{\sigma}$. Loading the sample from $\tilde{\sigma}$  to  
$\tilde{\sigma}+d\tilde{\sigma}$   the strain increment $d\tilde{\epsilon}$ is obtained. 
Then the sample is unloaded, back to the original $\tilde{\sigma}$, and one finds a remaining 
strain $d\tilde{\epsilon}^p$,  that corresponds to the plastic  component of the incremental 
strain. For small stress increments the unloaded stress-strain path is almost elastic. 
Thus, the difference   $d\tilde{\epsilon}^e = d\tilde{\epsilon}- d\tilde{\epsilon}^p$ 
can be taken as the elastic component of the strain.   

This procedure is implemented on one sample choosing different stress directions. 
Fig.\ \ref{de1} shows the load-unload stress paths and the corresponding strain response
when an initial stress state with $q=p/4$ is chosen, corresponding to $\sigma_1=5p/4$ and 
$\sigma_3=3p/4$, a larger stress in vertical direction. The end of the load paths in the 
stress space maps into a strain envelope response $d\tilde{\epsilon}(\theta)$ in the 
strain space. Likewise, the end of the unload paths map into a plastic envelope  response 
$d\tilde{\epsilon}^p(\theta)$. The {\it yield direction} 
$\phi$ can be found from this response, as the direction in the stress space where the plastic 
response is maximal. This is close to $\theta=90^o$ in this case. The {\it flow direction} $\psi$
is given by the direction of the maximal plastic response in the strain space, which is close to 
$70^o$ in this example. We found that these directions do not agree, corresponding to a 
{\it non-associated flow rule} as it is observed in experiments on realistic soils 
\cite{tatsuoka}. If one approximates the  plastic envelope response by its projection in the 
flow direction, the elastic response can be written as the simple form

\begin{equation}
d\tilde{\epsilon}^p=\frac{ \langle \hat{\phi} \cdot d\tilde{\sigma} \rangle }{h}\hat{\psi}.
\label{plastic}
\end{equation}

The so-called {\it hardening modulus} $h$  is the ratio between the modulus of the
maximal plastic strain and the modulus of the incremental stress. The function
$\langle x \rangle$ is defined as zero if $x\le 0$;  Otherwise it is valued to $x$. 
The unit vectors
$\hat{\psi}$ and $\hat{\phi}$ define the flow direction and the yield direction,
respectively. The vector  $ d\tilde{\sigma}$ defines the direction and magnitude of 
applied load. 

\end{multicols}

\begin{figure}
 \begin{center}
 \epsfig{file=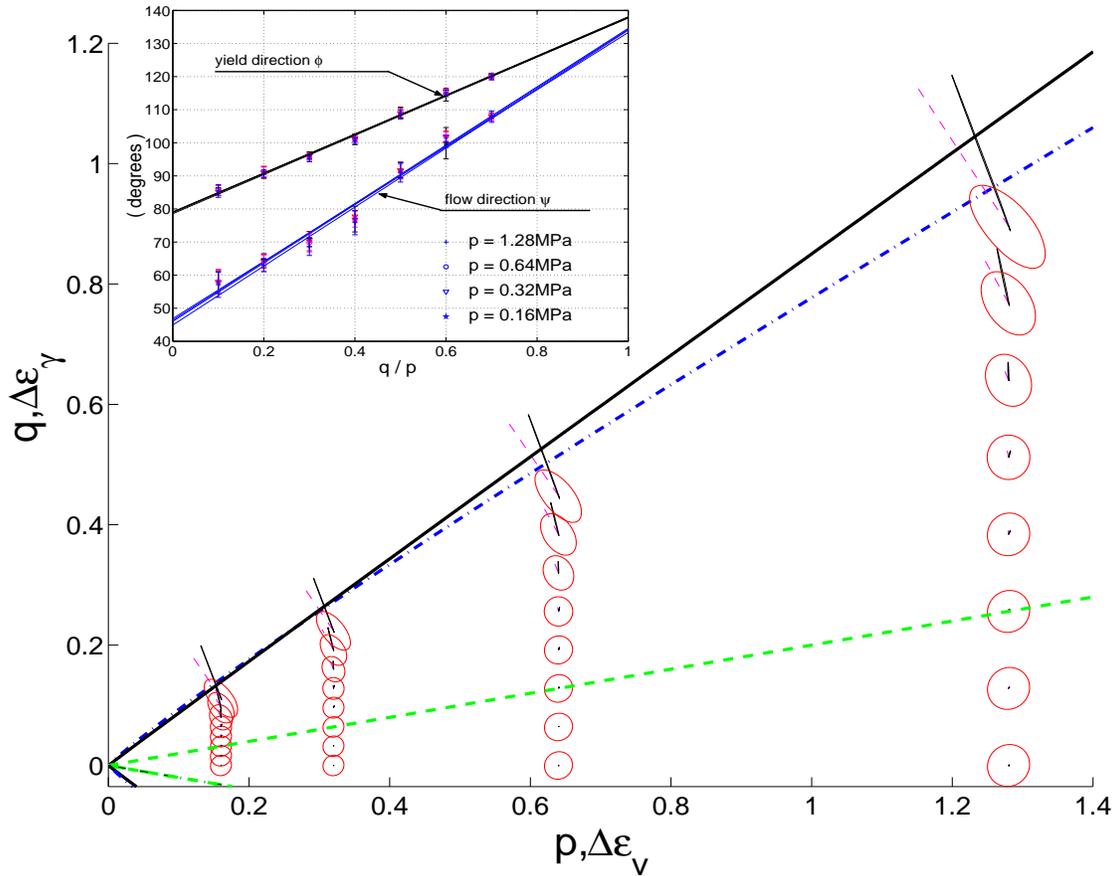,width=15.0cm,angle=0,clip=1}
 \end{center}
 \caption{
Elastic response $d\tilde{\epsilon}^e$ and plastic response $d\tilde{\epsilon}^p$ for different
levels of pressure $p$ and shear stress $q$. The envelope responses result
from the application of different loading modes with $|d\tilde{\sigma}|=10^{-4} p$. 
The elastic response, calculated  from Eq.\ (\ref{elastoplastic}), has a centered ellipse as 
envelope response. The plastic strain response lies almost on a straight line oriented in the
flow direction. This direction does not correspond with the yield direction (dashed lines).
The dash-dotted line represents the failure surface. When a stress value
above this line is applied the system fails. The solid line represents
the plastic limit surface.  The latter is obtained connecting the points where the plastic
deformation diverges. (That means $d\tilde{\epsilon}^p/|d\tilde{\sigma}| \rightarrow \infty$) 
The dashed line $q=0.18p$ represents the limit of the zone where the isotropic elasticity assumption 
is valid. The plot shows the yield and the flow directions as a function of 
the deviatoric stress ratio $q/p$, evaluated from the average over five different samples.
These calculations are made with fixed $k_n=160$\,MPa. }
 \label{elpl}
\end{figure}

\begin{multicols}{2}

Both elastic and plastic envelope responses are calculated from different stress values. 
The results are shown in  Fig. \ref{elpl}. The elastic response, calculated  
from Eq.\ (\ref{elastoplastic}), has a centered ellipse as envelope response for all 
the cases. Since the direction of this response does not always correspond to 
the direction of the volumetric strain increment, the general form for the elastic 
stiffness must be written as

\begin{equation}
d\tilde{\epsilon}^e=\frac{2}{E}\left[ \begin{array}{cc}
                  1-\nu&   -\alpha  \\
                  -\alpha  &   1+\nu  
\end{array} \right]d\tilde{\sigma}.
\label{elastic}
\end{equation}

\noindent
Here  $E$ and $\nu$ are the  classical parameters of the elasticity, i.e. the Young modulus  
and the Poisson ratio. They are not material parameters because they depend on the stress 
state we take. Moreover, an additional variable $\alpha$ must be included in this relation, 
taking into account the anisotropy of the elastic response. A limit of isotropy is
found around  $q = 0.18p$ (See Fig.\ \ref{elpl}). Below of this line the parameters 
of elasticity are almost
constants with $\alpha \approx 0$. Above this limit the stiffness decreases as a result of
open contacts, giving rise to an anisotropic elastic response.

In a previous work \cite{alonso}, the elasto-plastic quantities resulting form Eq.\ 
(\ref{plastic}) and Eq.\ (\ref{elastic}) have been evaluated as a function of the stress state. 
Since the mechanical response of soils depends not only on the initial stress state but also 
on the way how this state is reached \cite{diafalas}, these results are only valid in the 
case of monotonic load. In the classical theory of elasto-plasticity, the 
dependence of the strain response on the history of the deformation
is described by the evolution of the so-called 
{\it yield surface}. This surface encloses a hypothetical region in the stress space where 
only elastic deformations are possible \cite{drucker}.

\begin{figure}[htb]
 \begin{center}
 \epsfig{file=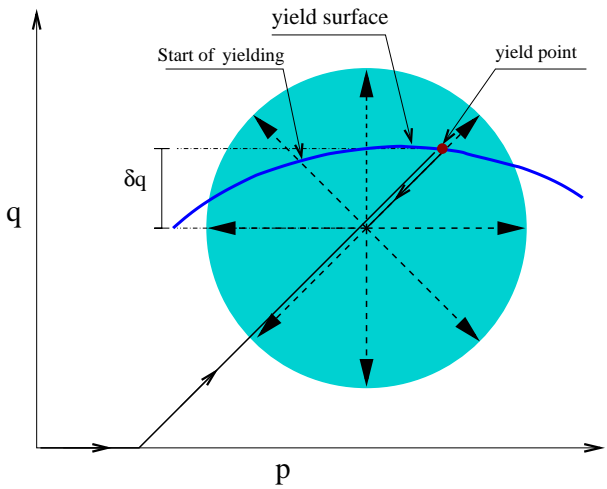,width=8.0cm,angle=0,clip=1}
 \end{center}
 \caption{Experimental procedure used to obtain the yield surface. Load-unload-reload tests
are performed, and the points in the reload path, where the yielding begins are marked. The
yield function is constructed by connecting these points. }
 \label{yield}
\end{figure}

We attempt to detect the yield surface by using a standard procedure proposed in experiments  
with sand \cite{tatsuoka}. Fig.\ \ref{yield} shows this procedure: initially  the sample is 
subject to isotropic pressure. Then the sample is loaded in the axial direction until it 
reaches a the yield-stress state with pressure $p$ and shear stress $q$. 
Since plastic deformation is found in this stress value, the point $(p,q)$ can be considered as
a classical yield point. Then, the classical theory assumes the existence of a yield surface 
containing this point. In order to explore the yield surface, the sample 
is unloaded in the axial direction until it reaches the stress point with pressure $p-\delta p $ and 
shear stress $q-\delta p$ inside  the elastic regime. Then the yield surface is constructed by
taking different directions in the stress space for re-loading. In each direction, the new yield 
point must be detected by a sharp change in the slope in the stress-strain curve, indicating
plastic deformations.

\begin{figure}[htb]
 \begin{center}
 \epsfig{file=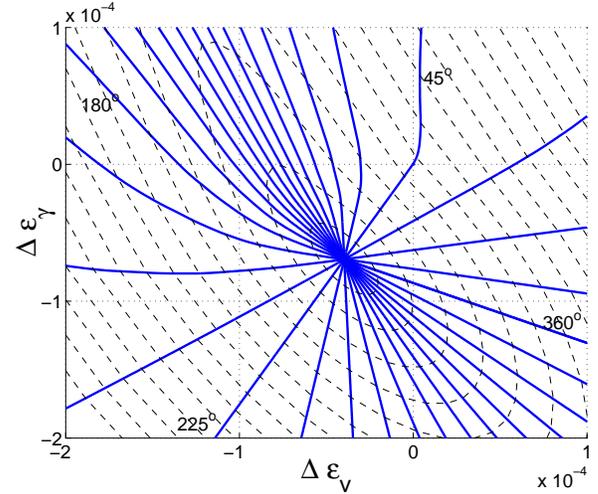,width=9.0cm,angle=0,clip=1}
 \end{center}
\caption{Strain response resulting from the preparation of the sample according to
Fig.\ \ref{yield}. The solid lines show the strain response for different loading directions.
The dashed contours connect the strain  increments values with the same value of
$|\Delta\tilde{\sigma}|$. The Figure shows that in any reload path different to that one
of the original load the yielding develops continuously. Thus, it is  not possible to
distinguish an elastic regime.}
\label{De}
\end{figure}

Fig.\ \ref{De} shows the strain response taking different load directions in the same sample.
If the direction of the reload path is the same as that one of the original load ($45^o$),  
we observe a sharp decrease of stiffness when the load point reaches the initial yield point,
which corresponds to the origin in Fig.\ \ref{De}.
However, if we take a different direction of re-loading, we find that the decrease of the 
stiffness with the loading becomes smooth. Since there is no  straightforward way to identify those 
points where the yielding begins, the yield function, as it was introduced by Drucker \& Prager
\cite{drucker} in order to describe a  sharp transition between the elastic and plastic regions 
is not consistent with our results.

\section{concluding remarks}
\label{conclusion}
The incremental elasto-plastic response of a Voronoi tessellated sample of polygons has been examined 
in the framework of the classical theory. The resulting constitutive relation leads to non-linear,
anisotropic elasticity, where the classical parameters of elasticity, the Young modulus and the 
Poisson ratio, are not material constants. The plastic response reflects the non-associated 
features of  realistic soils.  Here the classical analysis of Drucker \& Prager is not applicable,
because it is not possible to determine an elastic regime. 

Future work will be oriented to a micro-mechanically based description of these elasto-plastic
features. Since the mechanical response of the granular sample is represented as a collective 
response of all the contacts, it is expected that the macro-mechanical response can be completely 
characterized by the inclusion of some field-variables, which contain the information about the 
micro-structural arrangements  between the grains. Some statistical variables like the
fabric tensor have been included as internal variables \cite{thornton,mark}. This description, 
however, does not seem to offer a complete characterization of the constitutive response. 
More salient aspects, such
as the yielding of the contacts, and the fluctuations of the stress inside the granular material,
might offer a more complete set of internal variables for the description of the macroscopic state.
This is, in our point of view, an important challenge in the future.

\section*{Acknowledgments}

We thank F. Darve, P. Vermeer, F. Kun, and J. \AA str\o m for helpful discussions
and acknowledge the support of the {\it Deutsche Forschungsgemeinschaft\/} within the 
research group {\it Modellierung koh\"asiver Reibungsmaterialen\/}.

\end{multicols}


\begin{references}

\bibitem{vermeer} P. A. Vermeer, 
{\sl "A five-constant model unifying well-established concepts"\/}
in {\sl Constitutive Relations of soils}, edited by G. Gudehus, F. Darve and
L. Vardoulakis, Balkema Rotterdam, 175-197 (1984) 

\bibitem{darve}  F. Darve, {\sl "Incremental non-lineal constitutive relationships"\/}
in {\sl Geomaterials Constitutive Equations and Modeling\/}, 
edited by F. Darve, Elsevier applied Science, London 123-148 (1990)

\bibitem{kolymbas} D. Kolymbas 
{\sl "An outline of hypoplasticity"} 
Arch. App. Mech. {\bf 61\/}, 143-154 (1991)

\bibitem{cundall} P.A Cundall,
{\sl "Numerical experiments on localization in frictional materials"  \/} 
Ingenieur-Archiv, {\bf 59\/}  879-908 (1989)

\bibitem{radjai} F. Radjai, M. Jean, J.J. Moreau and S. Roux 
{\sl "Force distributions in Dense Two Granular systems"}
Phys. Rev. Lett. {\bf 77\/}, 2  274-277 (1996)

\bibitem{thornton} C. Thornton {\sl Computer simulated deformation of compact granular assemblies\/} 
Acta Mechanica {\bf 64\/}, 45-61 (1986)

\bibitem{bardet} J.P. Bardet, 
{\sl "Numerical solutions of incremental response of idealized granular materials"\/} 
Int. J. Plasticity {\bf 10\/}(8)  879-908 (1994)

\bibitem{tillemans} H. Tillemans and H.J. Herrmann,
{\sl "Simulating deformations of granular solid under shear"\/}
Phys. A {\bf 217\/},  261-288 (1995)

\bibitem{Kun} F. Kun and  H.J. Herrmann,
{\sl "A study of fragmentation process using a discrete element method"\/}
Comput. Methods Appl. Mech. Engrg. {\bf 138\/},  3-18 (1996)

\bibitem{addetta}G. A. Addetta, F. Kun, E. Ramm,
{\sl "On the application of a discrete model to the fracture process
of cohesive granular materials"\/}
Granular Matter {\bf 4}, 77-90 (2002)

\bibitem{superkun} F. Kun and H. J. Herrmann, {\it Fragmentation of Colliding Discs}, 
Int. Jour. Mod. Phys. C {\bf 7}, pp. 837-855 (1996).

\bibitem{Kun2} F. Kun and H. J. Herrmann, {\sl "Transition from damage to
  fragmentation in collision of solids"}, Phys. Rev. E {\bf 59}, 2623
(1999).
 
\bibitem{alonso} F. Alonso-Marroquin and H. J. Herrmann, {\sl
"Calculation of the incremental stress-strain relation of a polygonal packing"  }, 
Phys. Rev. E (2002) in press.

\bibitem{bagi}K. Bagi, {\sl" Microstructural Stress Tensor of Granular Assemblies with Volume Forces"}
Journal of Applied Mechanics {\bf 66}, 4 934-936

\bibitem{kruyt} N.P. Kruyt and L. Rothenburg, 
{\sl "Micro-mechanical Definition of the strain tensor for granular materials"\/}
ASME Journal of Applied Mechanics {\bf 118\/},  706-719 (1996)


\bibitem{diafalas} Y. F. Diafalas and E. P. Popov. {\sl "Plastic Internal Variables formalism of
Cyclic Plasticity"} Journal of Appliedd Mechanics {\bf 43} 645-650 (1976)

\bibitem{tatsuoka} F. Tatsuoka and K. Ishihara, {\sl"Yielding of sand in triaxial compression"} 
Soil and Foundations {\bf 14 \/}, 2  63-76 (1974)

\bibitem{drucker} D.C. Drucker and W. Prager.
 {\sl" Soil Mechanics and plastic analysis of limit design"},
Q. Applied Math.{\bf 10\/}(2) 157-165 (1952)

\bibitem{mark} M. L\"atzel, S. Luding, and H. J. Herrmann, 
{\sl "Macroscopic material properties from quasi-static, microscopic simulations of a two-dimensional shear-cell"\/} Granular Matter {\bf 2\/}(3), 123-135, (2000) 

\end{references}
\end{document}